# Stock Volatility Prediction using Time Series and Deep Learning Approach

Ananda Chatterjee
*Department of Data Science*
*Praxis Business School*
Kolkata, India
ananda.chatterjee89@gmail.com

Hrisav Bhowmick
*Department of Data Science*
*Praxis Business School*
Kolkata, India
hrisavbhowmick@gmail.com

Jaydip Sen
*Department of Data Science*
*Praxis Business School*
Kolkata, India
jaydip.sen@acm.org

*Abstract*—Volatility clustering is a crucial property that has a substantial impact on stock market patterns. Nonetheless, developing robust models for accurately predicting future stock price volatility is a difficult research topic. For predicting the volatility of three equities listed on India's national stock market (NSE), we propose multiple volatility models depending on the *generalized autoregressive conditional heteroscedasticity* (GARCH), *Glosten-Jagannathan-GARCH* (GJR-GARCH), *Exponential general autoregressive conditional heteroskedastic* (EGARCH), and LSTM framework. Sector-wise stocks have been chosen in our study. The sectors which have been considered are *banking*, *information technology* (IT), and *pharma*. yahoo finance has been used to obtain stock price data from Jan 2017 to Dec 2021. Among the pulled-out records, the data from Jan 2017 to Dec 2020 have been taken for training, and data from 2021 have been chosen for testing our models. The performance of predicting the volatility of stocks of three sectors has been evaluated by implementing three different types of GARCH models as well as by the LSTM model are compared. It has been observed the LSTM performed better in predicting volatility in *pharma* over *banking* and *IT* sectors. In tandem, it was also observed that E-GARCH performed better in the case of the *banking* sector and for *IT* and *pharma*, GJR-GARCH performed better.

## I. INTRODUCTION

Whereas developing robust models to make accurate future market predictions has been a major topic of research, assessing and forecasting future stock volatility is more difficult. Surprisingly, while stock market volatility can pose a considerable menace to shareholders if properly anticipated and controlled, it can also be a source of significant financial returns. Even when stock markets oscillate, fall, or skyrocket, there is always a profit window if market volatility is exploited. Volatility is traditionally defined as the dispersion of a stock's return series, as computed by taking the standard deviation of the return series. The standard deviation of a stock's return series indicates how tightly the stock's return values are concentrated around its mean value. The volatility of the series is modest if the return values are strongly connected across time. A high standard deviation, on the other hand, indicates that the series is highly volatile.

In a recent study, high variations in stock prices were predicted using a deep learning-based LSTM model basis on the data of 2020 when the variations of stock prices were too high [1].

While many factors influence stock market volatility, geographical and economic aspects like rules of taxation, monetary and revenue schemes, as well as interest rates, have considerable aftermath on market directional shifts, impacting volatility to a significant degree. When the Reserve Bank of India, for example, circulates an alteration in the rates of short-term borrowing for banks, the stock market immediately becomes more volatile. Bear market volatility causes investors to rebalance their stock holdings. However, accurately projecting future stock market volatility is a difficult endeavor. Even though there are numerous postulations and models in different publications, the majority of which have been proven to perform poorly in real-world applications, resulting in erroneous volatility forecasting.

A novel method was proposed by a few researchers to predict the stock price amidst volatility of the market based on the idea of a *State Frequency Memory* (SFM) recurrent network that can capture multi-frequency trading patterns from historical market data and make long and short-term predictions over time [2].

An ARIMA-based model was developed to predict Stock market volatility in a work [3]. A *support vector regression* (SVR) based approach was taken for stock price prediction in three different markets [4]. In another study, a mix of two approaches *self-organizing map* (SOM) neural network and *genetic programming* (GP) were adopted as SOM-GP to predict the stock price in highly volatile situations and it succeeded in predicting where the daily closing price was found to alternatively rise and fall [5].

This paper is divided into five sections. Section II defines the problem statement that this study will attempt to solve. In Section III, we take a look at some associated stock price volatility clustering. In Section IV, the methods used to solve the problem are described. In Section V, the experimental results of various methodologies are presented, and finally, in Section VI, the paper is concluded with the identification of some prospective future research directions.

## II. PROBLEM STATEMENT

This study aims to develop a reliable model for predicting stock price volatility. Three different sectors *banking, IT*, and *pharma* were chosen and the stock prices of these three sectors were considered to train the models picked up for this investigation to forecast volatility. Three sets of GARCH models were constructed, tuned up, and validated using stock prices of these sectors, and in tandem, the performance of three GARCH models on these sectors was compared. Apart from that, an LSTM-based approach was also taken for stock volatility prediction and the performance of that model was observed in the three considered sectors. Following that, a comparative analysis was made among the set of GARCH models and the LSTM in terms of their performance in predicting stock price volatility in *banking*, *IT*, and *pharma* domains.

### III. RELATED WORK

Researchers have done substantial work to predict stock volatility on the Ghana Stock Exchange using *random walk*, GARCH, and TGARCH models [6]. In another set of research, a group of GARCH models was used to predict the volatilities of stocks from several sectors listed in the NSE of India [7, 8]. In another study, stock volatility was forecasted in Nigerian Stock Market using a set of GARCH models, using the *normal*, *Student's t-distribution*. The generalized error distributions were studied to identify the best model for predicting future volatilities [9]. EGARCH, GJR-GARCH, and APARCH models were used in the analysis of volatility using Aviv Stock Exchange Indices in a study where EGARCH with student *t*-distribution came out as the most successful model for forecasting volatility [10]. A Researcher predicted the volatility of the SSE composite index using GARCH, TGARCH, and EGARCH models [11]. The GARCH-MIDAS model was used to predict symmetric and asymmetric volatility of stock price by a few researchers [12]. In subsequent work, an improved version of the GARCH-MIDAS model has been proposed [13]. In another work, it was observed that the study of the distribution of the residuals of GARCH (1, 1) models enables a more accurate assessment of the volatilities of the series [14]. In a study, LSTM performed better compared to GARCH in predicting stock price index volatility [15]. In another study, a model was proposed based on the CNN-LSTM combination to predict the volatility in Gold's price [16]. A comparative study was published in terms of stock volatility prediction using three neural network models such as RNN, DNN, and LSTM, and it was observed that RNN LSTM outperformed DNN [17]. An LSTM model was proposed to predict the volatility of specific factors of stock [18]. An RNN and LSTM-based approach was applied to predict the volatility of two financial stock indexes S&P500 and AAPL. Along this GARCH model was also used to predict the same, but LSTM outperformed GARCH [19].

### IV. METHODOLOGY

In this work, three sectors were chosen from those stock prices collected to forecast the stock volatility. The sectors are *IT*, *banking*, and *pharma*. The data was extracted from Yahoo Finance website. The period for which data was extracted from these sectors lies between 2017 to 2021. The data collected from January 2017 to December 2020 was utilized for training, while the entire data set from 2021 was used for testing. The data had 5 features such as *'High,' Low', 'Open', 'Close', and 'Adj. Close'*, among these *'Close'* columns, was used. A gamut of Time Series and Deep Learning-based models were used in our study to predict volatility. Among the time series models, a set of GARCH models (GARCH, EGARCH, and GJRGARCH) were used and from a deep learning-based approach LSTM was used. All of these four models were trained on the dataset of the period Jan 2017 to Dec 2020 and tested on the dataset for the whole of 2021. The working principle and model-building approach of these four models have been discussed below.

#### A. Volatility Calculation

To compute the actual volatility present in the data firstly the *'Return'* was calculated by taking the percentage change on the *'Close'* column using the *pct_change* function, then multiplying it by 100. Thereafter daily volatility was calculated by taking the standard deviation of the *'Return'* column using the *std* function. The monthly volatility was calculated by multiplying the daily volatility with sqrt (21) and the annual volatility was calculated multiplied by the daily volatility with sqrt (252).

#### B. GARCH

Before applying GARCH the order of *p*, *d*, and *q* was determined using the *auto_arima* function where *p*, *d*, and *q* where *p* refers to the Autoregressive model lags, *d* is the differencing needed to achieve a stationary series, and *q* refers to the Moving Average lags. The optimum values of *p*, *d*, and *q* came out as $p = 1$, $d = 0$, and $q = 1$. The optimum value of these values was chosen seeing the *Bayesian information criteria* (BIC) value, for which this BIC is the lowest that combination was chosen as optimum. Thereafter the GARCH model was fit using the *arch_model* function with a combination of *p = 1, q = 1, mean='constant', and dist='skewt'*. Here, *skewed-t* distributed residuals were used over normally distributed residuals as stock values follow this over normal distribution. A *t-distribution* is comparable to a normal distribution, except it has a larger kurtosis and flatter tails that encompass fewer extreme values. Furthermore, the volatility of the GARCH (1, 1) model was displayed against the stock's daily return using a *skewed-t* error distribution to see how well the GARCH model captures the daily return series' pattern of fluctuations. The *summary* function was used to obtain the coefficient values of *omega, alpha, beta,* and *gamma* and their respective *p values*, and it was found that all the *p* values are smaller than 0.05. It indicates that the residuals fitted well into the model. The GARCH models fitted very well into the daily return values of the IT sector. However, the model fit was not very good on the *pharma* sector data.

#### C. GJR-GARCH

The GARCH model captures the volatility well depending on the length of the shock in stock prices, it doesn't depend on the sign of the shock, but there are instances where the sign causes a good impact on volatility like GARCH doesn't take care of if the shock causes the volatility is due to good or bad news, in those cases two asymmetric models GJR-GARCH and E-GARCH come into play.

The GJR-GARCH model was fit via means of the *arch_model* function using the following parameters: *p = 1, q = 1, o = 1, mean = 'constant', vol = 'GARCH',* and *dist = 't'* which created a GJR-GARCH model with constant mean and residuals are following skewed *t* distribution. Using the *summary* function the coefficient values of *omega, alpha, beta, gamma*, and the *p values* that go with them were obtained. From the *p values,* it was observed that for all the parameters it fell less than 0.05 which caused the goodness of fit of this model into residuals. GJR-GARCH performed best in all three sectors among three sets of GARCH models.

#### D. EGARCH

The EGARCH model was formed utilizing the *arch_model* function on the *Return* column which was computed using the percentage change of the *Close* column. The parameters used to build this model were *p = 1, q = 1, o*

= 1, mean = 'constant', vol = 'EGARCH', and dist = 't' which created a GJR-GARCH model with constant mean and residuals following *skewed t* distribution. Using the *summary* function the coefficient values of *omega*, *alpha*, *beta*, and *gamma*, with associated *p* values were obtained. From the *p* values, it was observed that for all the parameters it fell less than 0.05 which caused the goodness of fit of this model into residuals. Among the three sectors, EGARCH performed best in *IT* and poor in *pharma*. In both cases for *IT* and *banking*, it outperformed GARCH but in the case of *pharma*, GARCH performed better than EGARCH.

### E. LSTM

For deep learning, stacked LSTM was used for model building. The data from three sectors *banking*, *IT*, and *pharma* from the period 2017 to 2020 (980 records) was used as a training dataset, while the test dataset consisted of data for the whole period of 2021 (246 records). The *Return* column, which is nothing more than the percentage change of the *Close* column, was subjected to a univariate analysis. Following that, the data for training and testing was compiled. The time step utilized was a seven-day rolling window. The goal was to build *X_train* on the first five days of the train set (index: 0 to 4). The value on the 6$^{th}$ day (index: 5) will be *y_train* based on this. The sliding window advances by a day after that. The values stored in an array over the following five days (index: 1 to 5) will be in the *X_train*, and the value for the seventh day (index: 6) will be in the y-train. This will continue until the *X_train* has values up to the second last record in the train set, and the *y_train* has the train set's last index. The train set will last 974 days in all. After then, the counter is moved to the testing set. The *X_test* will be applied to the first seven days' value on the test set, while the *y_test* will be assigned to the eighth day's value. This will continue till the test set is completed.

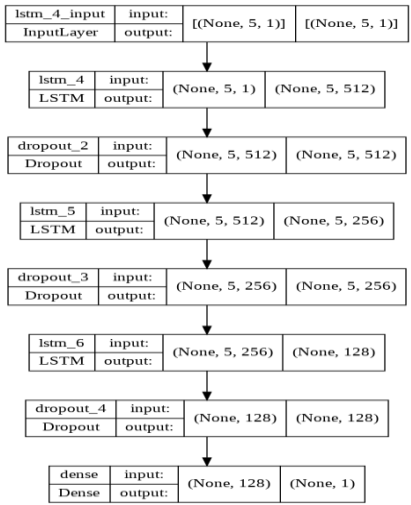

Fig. 1. The LSTM model for stock volatility prediction

An LSTM is made up of an RNN cell and three gates (forget, input, and output). The forget gate decides whether the incoming information is to be kept or removed from the memory. The input gate determines how much data is needed for the cell. Internal Cell State's output gate determines which outputs to keep. The number of time steps in the input layer of the LSTM model created for this study was set to five, and the number of features was set to one (which is the close column). The input layer of the LSTM has a shape of (974, 5), where 974 is the no. of observations in the training dataset and 5 is the no. of time steps considered. The first hidden layer was created using 512 nodes, which was followed by 20% dropout, 256 nodes were used to form the second hidden layer followed by 20% dropout, third hidden layer comprised 128 nodes, which was again followed by 20% dropout. *Rectified linear unit* (RELU) activation function was employed in the case of the input and three hidden layers. After creating all hidden layers, a dense layer was used at last along with the *linear* activation function. The model was compiled using the *Adam* optimizer and *mean squared error* (MAE) as the error function. For the LSTM model, the selected *batch size* was 64, and in the model, the number of epochs was chosen to be 100, for training the network.

## V. RESULTS

This section presents the results. For the three sectors, three GARCH models, the basic GRACH, and the two asymmetric GARCH models, GJR-GARCH and EGARCH, are built. Additionally, an LSTM model is also designed. Using these models, the volatilities of the daily return values of the index of the three sectors for the training and the test periods are evaluated. To evaluate the performance of the models, their RMSE values are computed and compared.

### A. Banking Sector

The daily return values are computed for the NSE *banking* sector daily index from January 3, 2017, to December 31, 2020, and the GARCH models are trained on the training data. The daily volatility values of the *banking* sector index for the test period from January 1, 2021, to December 31, 2021, are predicted using the models.

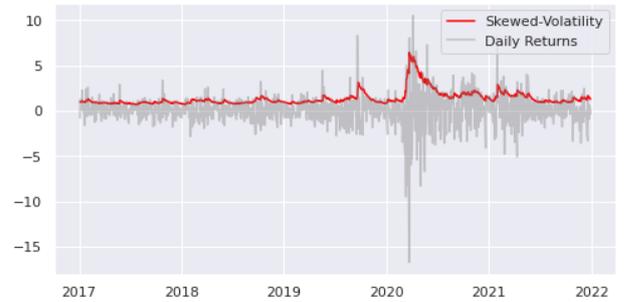

Fig. 2 The daily return vs the predicted return by the GARCH model having *skewed-t* distributed error for the banking sector index.

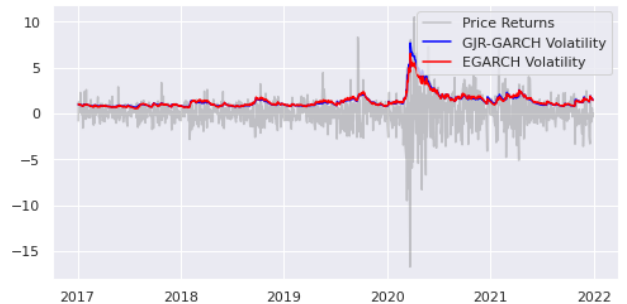

Fig. 3 The actual volatility of the daily return vs the volatility predicted by the GJR-GARCH and the EGARCH models for the banking sector index.

Figure 2 depicts the GARCH model predicted volatility with *skewed-t* distributed residuals vs. the actual daily return values for the NSE *banking* sector index for the entire period from January 2017 to December 2021.

Figure 3 exhibits the volatility of the daily return series of the NSE *banking* sector, the GJR-GARCH volatility, and the EGARCH volatility. It is evident, that the performances of both the GARCH models are almost identical, and both models have accurately captured the volatility of the daily return series of the *banking* sector's daily index.

Table I. PERFORMANCE OF THE GARCH MODELS FOR THE NSE BANKING SECTOR

| Models | Training Data | | Testing Data | |
|---|---|---|---|---|
| GARCH | RMSE | 10.1088 | RMSE | 10.0794 |
| GJR-GARCH | RMSE | 10.0248 | RMSE | 9.9970 |
| EGARCH | RMSE | 9.9467 | RMSE | 9.9195 |

To evaluate the performance of the three GARCH models on the daily return series of the NSE *banking* sector, the backtesting of the models is carried out over the training and the test data. Table I presents the *root mean square error* (RMSE) of the three models. A sliding window of 10 days is used in backtesting. While the RMSE values for all three models are quite low, the EGARCH model is found to be the most accurate GARCH model for the *banking* sector index.

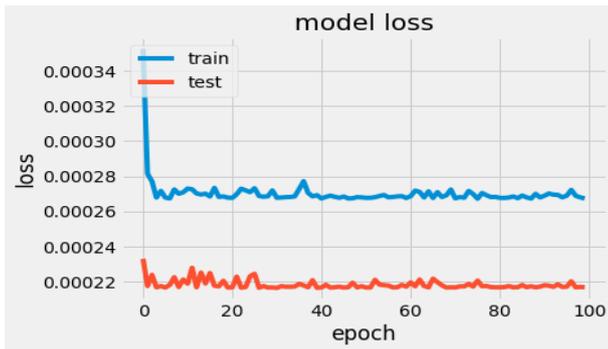

Fig. 4 The training vs test loss of the LSTM model plotted for different epoch values for the banking sector return series.

TABLE II. PERFORMANCE OF THE LSTM MODEL FOR THE BANKING SECTOR

| Model | Training Data | | Testing Data | |
|---|---|---|---|---|
| LSTM | RMSE | 0.0163 | RMSE | 0.0147 |

The training and the test loss for the LSTM model which was trained over 100 epochs are plotted in Figure 4. Table II presents the backtesting results of the LSTM for the training and the test data. The RMSE of the LSTM model for the training and the test data presented in Table II are found to be significantly lower than the corresponding values of the GARCH model presented in Table I.

*B. IT Sector*

The daily return values are computed for the NSE *information technology* (IT) sector daily index from January 3, 2017, to December 31, 2020, and the GARCH models are trained on the training data. The models are then used to predict daily volatility values of the IT sector index for the test period from January 1, 2021, to December 31, 2021.

Figure 5 depicts the GARCH model predicted volatility with *skewed-t* distributed residuals vs. the actual daily return values for the NSE IT sector index for the entire period from January 2017 to December 2021. It is observed that the GARCH model can capture the volatility pattern of the return series very accurately.

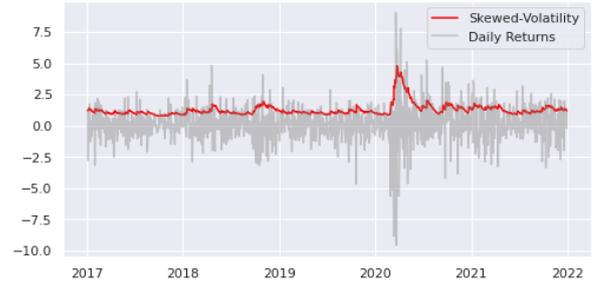

Fig. 5 The daily return vs the predicted return by the GARCH model having *skewed-t* distributed error for the IT sector index.

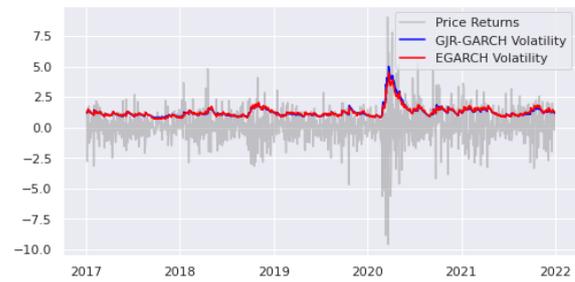

Fig. 6 The actual volatility of the daily return vs the volatility predicted by the GJR-GARCH and the EGARCH models for the IT sector index.

Figure 6 exhibits the volatility of the daily return series of the NSE IT sector, the GJR-GARCH volatility, and the EGARCH volatility. It is evident, that the performances of both the GARCH models are almost identical, and both the models have accurately captured the volatility of the daily return series of the IT sector daily index.

TABLE III. PERFORMANCE OF THE GARCH MODELS FOR THE NSE IT SECTOR

| Models | Training Data | | Testing Data | |
|---|---|---|---|---|
| GARCH | RMSE | 5.2964 | RMSE | 5.3217 |
| GJR-GARCH | RMSE | 5.2359 | RMSE | 5.4157 |
| EGARCH | RMSE | 5.2956 | RMSE | 5.3216 |

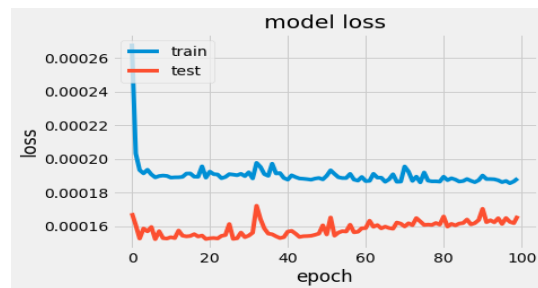

Fig. 7 The training vs test loss of the LSTM model plotted for different epoch values for the IT sector return series

The performance of GARCH models on the daily return series of the NSE IT sector is evaluated by backtesting of the models. Table III presents the *root mean square error*

(RMSE) of the three models. A sliding window of 10 days is used in backtesting. The RMSE values for all three models are found to be appreciably lower than those of the *banking* sector. The EGARCH model is again found to be the most accurate one as it has yielded the lowest RMSE on the test data.

The training and the test loss for the LSTM model which was trained over 100 epochs are plotted in Figure 7. Table IV presents the backtesting results of the LSTM for the training and the test data. The RMSE of the LSTM model for the training and the test data presented in Table IV are found to be significantly lower than the corresponding values of the GARCH model presented in Table III.

TABLE IV. PERFORMANCE OF THE LSTM MODEL FOR THE IT SECTOR

| Models | Training Data | | Testing Data | |
|---|---|---|---|---|
| LSTM | RMSE | 0.0138 | RMSE | 0.0125 |

### C. Pharma Sector

The daily return values are computed for the NSE *pharma* sector daily index from January 3, 2017, to December 31, 2020, and the GARCH models are trained on the training data. The models are then used to predict daily volatility values of the *pharma* sector index over the test period from January 1, 2021, to December 31, 2021.

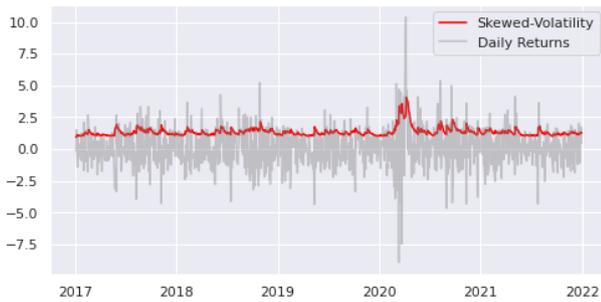

Fig. 8 The daily return vs the predicted return by the GARCH model having a *skewed-t* distributed error for the pharma sector index.

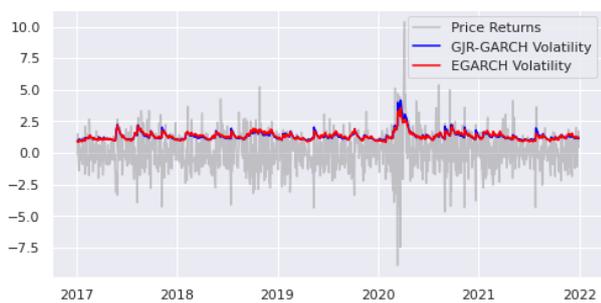

Fig. 9 The actual volatility of the daily return vs the volatility predicted by the GJR-GARCH and the EGARCH models for the pharma sector index.

Figure 8 shows the volatility predicted by the GARCH model having residuals following *skewed-t* distribution vs. the actual daily returns for the NSE *pharma* sector index for the entire period from January 2017 to December 2021. It is observed that the GARCH model has been able to accurately capture the volatility.

Figure 9 exhibits the volatility of the daily return values of the NSE *pharma* sector, the GJR-GARCH volatility, and the EGARCH volatility. It is evident, that the performances of both the GARCH models are quite similar. Both models are found to be accurate in capturing the volatility patterns of the *pharma* sector's daily returns.

TABLE V. PERFORMANCE OF THE LSTM MODEL FOR THE PHARMA SECTOR

| Models | Training Data | | Testing Data | |
|---|---|---|---|---|
| GARCH | RMSE | 4.9688 | RMSE | 4.6688 |
| GJR-GARCH | RMSE | 4.9448 | RMSE | 4.5432 |
| EGARCH | RMSE | 4.7648 | RMSE | 4.5245 |

For evaluating the performance of the three GARCH models on the NSE IT sector returns, the models are backtested on the training and the test data. Table V presents the RMSE values of the three models. A sliding window of 10 days is used in backtesting. The RMSE values for all three models are found to be the lowest among the three sectors. The EGARCH model is again found to be the most accurate one among the three GARCH models as it has yielded the lowest RMSE on both training and test data.

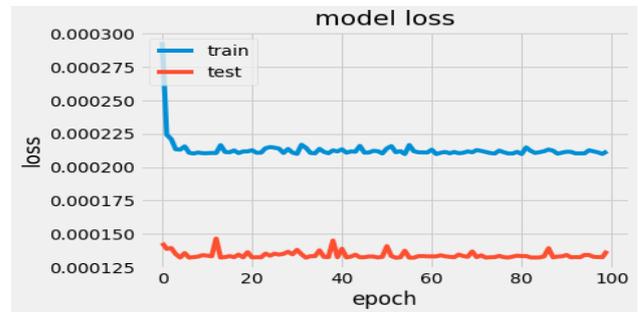

Fig. 10 The training vs test loss of the LSTM model plotted for different epoch values for the pharma sector return series

TABLE VI. PERFORMANCE OF THE LSTM MODEL FOR THE PHARMA SECTOR

| Models | Training Data | | Testing Data | |
|---|---|---|---|---|
| LSTM | RMSE | 0.0145 | RMSE | 0.0115 |

The loss exhibited by the LSTM model which was trained over 100 epochs is plotted in Figure 10. The backtesting results of the LSTM model are presented in Table IV. The RMSE of the LSTM model presented in Table VI is significantly low in comparison to the corresponding values of the GARCH models presented in Table IV.

### D. Comparative Study of GARCH models and LSTM on different sectors

Among the set of GARCH models, the performance of the E-GARCH on the test data is found to be superior to the other two models for all three sectors studied in this work. On the training data, EGARCH performed the best except for the IT sector. The performance of the GJR-GARCH model is found to be the best on the training data. The *Bayesian Information Criteria* (BIC) of a model is a good metric for model selection [20]. This metric makes a trade-off between its log-likelihood value (which shows how well the model has fitted into the training dataset), and its number of parameters. The BIC of a model is computed using (1).

$$BIC = -2 * LL + ln(N) * k \quad (1)$$

In (1), *ln* refers to the natural logarithm function, *LL* is the log-likelihood of the model, *N* is the size of the training dataset, and *K* is the number of parameters in the model. The model with the lowest BIC score is the most optimum one as it has been able to maximize the log-likelihood value using the minimum number of parameters among all the candidate models. The BIC values of the models for the three sectors are presented in Table VII.

TABLE VII. THE BIC SCORES OF THE GARCH MODELS

| Sectors | GARCH | GJR-GARCH | EGARCH |
|---|---|---|---|
| Banking | 3928.86 | 3908.07 | 3910.93 |
| IT | 3824.83 | 3822.76 | 3828.65 |
| Pharma | 4147.61 | 4133.71 | 4134.60 |

While the BIC values for the GJR-GARCH model are the minimum for all three sectors, the EGARCH models have the lowest RMSE. In the real world, financial time series exhibit asymmetric volatilities. EGARCH models can more effectively capture the asymmetric features in comparison to their GJR-GARCH, and hence, the performance of the EGARCH models is superior, in general [7].

The RMSE values of the models on the test data are summarized in Table VIII. The test period was from January 1, 2021, to December 31, 2021. The RMSE values reflect the root mean square error committed by the model in predicting the volatilities of the daily return values for the test period. For all three sectors, the LSTM model has been far more accurate in comparison to all three GARCH models. From Table VII, the EGARCH model is found to be the most accurate.

TABLE VIII. THE PERFORMANCE OF THE GARCH AND THE LSTM MODEL

| Stocks | Models | RMSE |
|---|---|---|
| Banking | GARCH | 10.0794 |
|  | GJR-GARCH | 9.9970 |
|  | EGARCH | 9.9195 |
|  | LSTM | 0.0147 |
| IT | GARCH | 5.3217 |
|  | GJR-GARCH | 5.4157 |
|  | EGARCH | 5.3216 |
|  | LSTM | 0.0125 |
| Pharma | GARCH | 4.6688 |
|  | GJR-GARCH | 4.5432 |
|  | EGARCH | 4.5245 |
|  | LSTM | 0.0115 |

## VI. CONCLUSION

In this work, four models have been presented for predicting the volatilities of return values of the daily index of three sectors of NSE. The three sectors studied in this work are *banking*, *information technology* (IT), and *pharma*. Three GARCH-based models including the basic GARCH, GJR-GARCH, and the EGARCH, and an LSTM model are proposed. The models are trained on the historical daily index values of the three sectors from January 1, 2017, to December 31, 2020. The testing of the models is carried out on the daily index values from January 1, 2021, to December 31, 2021. The models are compared on their RMSE values on the test data. It is observed that while the GJR-GARCH and the EGARCH have outperformed the basic GARCH, the performance of the EGARCH is superior to that of the GJR-GARCH models for all three sectors. However, the LSTM-based deep learning models have been far more accurate than the GARCH models as evident from their very low RMSE values for all three sectors. In future work, additional GARCH models, m-GARCH and o-GARCH, will be studied in other sectors of NSE.